\documentstyle[aps,eqsecnum,12pt]{revtex}
\begin{document}

\title{\bf No hair for spherical black holes:  charged and nonminimally
coupled scalar field with self--interaction}

\author{Avraham E. Mayo\thanks{Electronic mail: Mayo@venus.fiz.huji.ac.il}
and Jacob D. Bekenstein\thanks{Electronic mail: bekenste@vms.huji.ac.il}}

\address{\it The Racah Institute of Physics, Hebrew University
of Jerusalem,\\ Givat Ram, Jerusalem 91904, Israel}

\date{\today}

\maketitle

\begin{abstract}
gr-qc/9602057
We prove three theorems in general relativity which rule out classical
scalar hair of static, spherically symmetric, possibly electrically charged
black holes.  We first generalize Bekenstein's no--hair theorem for a
multiplet of minimally coupled real scalar fields with not necessarily
quadratic action to the case of a charged black hole.  We then use a
conformal map of the geometry to convert the problem of a charged (or
neutral) black hole with hair in the form of a neutral self--interacting
scalar field nonminimally coupled to gravity to the preceding problem, thus
establishing a no--hair theorem for the cases with nonminimal coupling
parameter $\xi<0$ or $\xi\geq {1\over 2}$.  The proof also makes use of a
causality requirement on the field configuration.   Finally, from the
required behavior of the fields at the horizon and infinity  we exclude hair
of a charged black hole in the form of a charged  self--interacting scalar
field nonminimally coupled to gravity  for any $\xi$.

\end{abstract}

\section{INTRODUCTION}

``Black holes have no hair'' was introduced by Wheeler in the early 1970's
as a principle predicting simplicity of the stationary black hole family.  The
proliferation in the 1990's of stationary black hole solutions with ``hair''
of various sorts \cite{hairy} may give the impression that the principle has
fallen by the wayside.  However, this is emphatically not the case for scalar
field hair, possibly accompanied by Abelian gauge fields.  The only
exceptions known to ``black holes have no hair'' in this department are the
Bronnikov--Melnikov--Bocharova--Bekenstein (BMBB) spherical extremal black
hole with electric charge and a scalar field nonminimally coupled to gravity
in conformally invariant fashion \cite{Bro70,Bek75},  its magnetic monopole
extension \cite{Virb}, and the Achucarro--Gregory--Kuijken (AGK) black hole
\cite{Achu}, a charged black hole transfixed by a Higgs local cosmic string.
Even these examples are not contrary to the spirit of the no--hair
conjecture: the first seems to be unstable \cite{Bro78}, the second is
too similar to the first to escape its fate, while the third is not
asymptotically flat.  What is the evidence for ``no scalar hair'' for black
holes ?

The first no--scalar hair theorems applied to the common massless
scalar field \cite{Cha,Bek72} and to the neutral Klein--Gordon field
\cite{Bek72,Teit}. The latter theorem's proof is also found to work for the
neutral scalar field with a monotonically increasing  self--interacting
potential.  Little progress was made in extending these theorems during the
1970's and 80's.  A notable exception was Adler's and Pearsons' theorem
\cite{Adl} which excludes charged Higgs hair for a charged black hole.  This
theorem has, however, occasionally been regarded as flawed \cite{Gib}.
Lately theorems by Heusler \cite{Heu}, Sudarsky \cite {Sud} and Bekenstein
\cite{Bek95} have become  available which exclude electrically neutral black
holes with hair as minimally coupled scalar fields endowed with positive
semidefinite self--interaction potentials of otherwise arbitrary shape.
The last mentioned theorem applies also to fields whose lagrangians are not
necessarily quadratic in the gradients of the fields.

Whereas simple scalar fields are covered by all these theorems, various
complications such as charge of the field and the hole, nonminimal coupling
to gravity, {\it etc.,\/} are not.  Early works in this more challenging
direction are the papers by Xanthopoulos and Zannias \cite{Xanthopoulos}
and  Zannias \cite{Zannias} which establish the uniqueness of the BMBB
black hole among the asymptotically flat static solutions of the Einstein and
{\it conformal\/} scalar field equations,  and the recent theorem by Saa
\cite{Saa} which excludes, for spherical black holes,  a broader, but still
limited, class of nonminimally coupled neutral scalar hair
(see Sec.\ref{map}).

In the present work we consider whether a charged black hole may possess hair
in the form of a scalar field with self--interaction and with nonminimal
coupling to gravity and gauge covariant coupling to the electromagnetic
field.  The motivation for looking at nonminimal gravitational coupling
is supplied by the existence of the BMBB black hole solution with
nonminimally coupled scalar hair.  The motivation for considering coupling of
the scalar to the electromagnetic field comes from the existence of the AGK
black hole.  Since nonminimal gravitational coupling entails not necessarily
positive field energy,  one looses one of the earlier tools for proving no
hair theorems
\cite{Bek95}.  Our assumption of spherical symmetry simplifies things enough
to allow us to prove several useful theorems.

In Sec.\ref{TBE} we formulate the equations of the scalar field
coupled nonminimally to gravity and gauge covariantly to the Maxwell field,
write down the energy--momentum tensor, and discuss restrictions on it from
regularity of the horizon and causality requirements.  The last, in
particular, do not seem to have been taken advantage of by previous workers.
Sec.\ref{GtCBH} generalizes a theorem by one of us \cite{Bek95} which
excludes hair in the form of a multiplet   of mutually interacting real
scalar fields with possibly nonquadratic kinetic action.  The theorem is here
extended to an electrically charged black hole, still under the assumption of
positivity of energy of the fields.  The extended theorem provides one of the
tools for proving, in Sec.\ref{xinontrivial}, a theorem ruling out, for an
electrically charged or neutral black hole, hair in the form of a neutral
scalar field with standard quadratic kinetic action, a positive semidefinite
self--interacting potential  and nonminimal coupling to gravity.  The theorem
is proved for the ranges
$\xi<0$ or $\xi\geq  {1\over 2}$ of the nonminimal
coupling parameter; its proof uses  a conformal map to convert the problem to
the one dealt with by the theorem of Sec.\ref{GtCBH}.  Also central to its
proof are the causality restrictions on the energy--momentum tensor.  In
Sec.\ref{ChargedField} a theorem is proved which rules out, for an
electrically charged black hole, hair in the form of a charged scalar field
with standard kinetic action, a positive semidefinite self--interaction
potential and nonminimal coupling to gravity with any $\xi$.  The proof,
which is given separately for nonextremal and
extremal black holes, centers on the behavior of the various fields
near the horizon and at infinity. Sec.\ref{conclusions} summarizes
our findings and speculates on their implications.

\section{BASIC EQUATIONS AND PHYSICAL RESTRICTIONS} \label{TBE}

Here we derive the energy-momentum tensor from the Maxwell-charged scalar
action with self--interaction and non-minimal coupling to gravity. Then we
derive the field equations for the scalar and the Maxwell fields.  Throughout
we use units with $c=1$.

\subsection{The Energy-Momentum Tensor}	
\label{TEMT}

We assume the existence of an asymptotically flat joint solution of the
 Einstein, scalar field and Maxwell equations, having the character of a
static, spherically symmetric, charged black hole spacetime. The symmetries
entitle us to write the metric outside the horizon as
\begin{equation}
	ds^2 = - e^\nu dt^2 + e^\lambda dr^2 + r^2 (d\theta^2 + \sin^2\theta\,
d\phi^2)
\label{metric}
\end{equation}
with $\nu(r)$ and $\lambda(r)$ both nonnegative and obeying $\nu(r),
\lambda(r) \sim O(r^{-1})$ as $r\rightarrow\infty$ because of asymptotic
flatness. The event horizon is at $r=r_H$ where $e^{-\lambda(r_H)}=0$ (see
Sec.\ref{general} below).  In case there are several such zeroes, the horizon
corresponds to the outer one. Anticipating the results of
Secs.\ref{Nonextremal} and \ref{extremal}, we note that near the event horizon
of a black hole of nonextremal or extremal kind
\begin{equation} e^\nu \sim
e^{-\lambda}\sim\cases{ r-r_H,&nonextremal;\cr (r-r_H)^2, &extremal.\cr}
\label{behavior_horizon}
\end{equation}
These results apply whatever the matter content of the spacetime.

The action of a charged scalar field with non-minimal coupling to gravity,
gauge covariant coupling to electromagnetism (or any $U(1)$ gauge field) and
with a general self--interaction potential is
\begin{equation}
	S_{SM\xi}=-{1\over 2}\int{\left(
		   D_\alpha \psi {(D^\alpha\psi)}^*
		  +  \xi R \psi \psi^*
	 	  +  V(\psi \psi^*)
		  + {1\over 8 \pi} F^{\alpha \beta} F_{\alpha \beta}
				\right) \sqrt{-g} d^4x}
\label{L}
\end{equation}
where $\psi$ is the complex scalar field, $A_{\mu}$ the Maxwell
vector potential, $D_{\mu}=\partial_{\mu}-iqA_{\mu}$ the gauge covariant
derivative ($q$ is the charge), $F_{\nu\mu}=A_{\mu,\nu}-A_{\nu,\mu}$  the
Faraday field tensor,  $V=V(|\psi|^2)$ the self--interaction potential,
$R$  the scalar curvature, and $\xi$  the strength of the nonminimal
coupling to gravity.  We assume throughout that $V$ is everywhere regular (V
and its first derivative bounded for finite argument) as well as positive
semidefinite.

The energy--momentum tensor that follows from the action is
\begin{eqnarray}
T_{\mu \nu}&=&{1\over 2} D_\mu \psi (D_\nu \psi)^*
		      + {1\over 2} (D_\mu \psi)^* D_\nu \psi
		      - {1\over 2} D_\alpha \psi (D^\alpha \psi)^*
			  g_{\mu \nu}
\nonumber \\
		      &-& \xi (\psi^* \psi)_{,\mu;\nu}
		       + \xi \hbox{\rlap{$\sqcap$}$\sqcup$} (\psi^* \psi) g_{\mu \nu}
		       + \xi (\psi^* \psi) G_{\mu \nu}
		       - {1\over 2}V g_{\mu \nu}
		       + T^{(em)}_{\mu \nu}
\label{TG}
\end{eqnarray}
where $G_{\mu \nu}$ denotes the Einstein tensor and
\begin{equation}
T^{(em)}_{\mu \nu}={1\over 4\pi}
	  \left(F_{\mu \alpha} F_\nu{}^\alpha
		-{1\over 4} g_{\mu \nu} F^{\alpha \beta} F_{\alpha \beta}
 		 \right)
\label{TEM}
\end{equation}
Here and elsewhere $\hbox{\rlap{$\sqcap$}$\sqcup$}$ is the d'Alembertian. By
virtue of the symmetries, the components of the electromagnetic
energy--momentum tensor satisfy \cite{Bek71}  \begin{equation}
	{T^{(em)}}^r_r={T^{(em)}}^t_t
		      =-{T^{(em)}}^\theta_\theta=-{T^{(em)}}^\phi_\phi
		      =-{Q^2\over 8\pi r^4}
\label{Tem}
\end{equation}
where $Q(r)$ is the electric charge enclosed by the sphere of radius $r$.
Eliminating $R$ from Eq.(\ref{TG}) by means of the trace of Einstein's
equations, $R=-8\pi GT$, we obtain
\begin{eqnarray}
	T^\nu_\mu (1-8\pi G \xi \psi^* \psi) &=&
			 {1\over 2} D_\mu \psi (D^\nu \psi)^*
		       + {1\over 2} (D_\mu \psi)^* D^\nu \psi
		       - {1\over 2} D_\alpha \psi (D^\alpha \psi)^*
		           \delta^\nu_\mu	
\nonumber \\
		      &-& \xi (\psi^* \psi)^{;\nu}_{,\mu}
		      + \xi \hbox{\rlap{$\sqcap$}$\sqcup$} (\psi^* \psi)
\delta ^\nu_\mu
		      - {1\over 2}V \delta^\nu_\mu + {T^{(em)}}^\nu_\mu
\end{eqnarray}

In light of the angular and temporal symmetries, $F_{tr}$ is the only
nonvanishing field component.  Thus only $A_r$ and $A_t$ need be
nonvanishing. Then the gauge transformation $A_\mu\rightarrow A_\mu
+\Lambda_{,\mu}$ with $\Lambda = -\int A_r\, dr$ removes $A_r$.  The new
$A_t$ must have the form $f(r)+g(t)$ in order to give a stationary $F_{tr}$.
A further gauge transformation with $\Lambda =-\int g(t)\, dt$ makes $A_t$
static.  In this second
 gauge any temporal variation of the phase of $\psi$ must be linear
in $t$ in order that both the current and charge density be time independent.
More precisely, the phase must be $\varphi(r)-\omega t$ with $\omega$ a real
constant. A last gauge transformation with $\Lambda=-\omega t/q$
reduces $\psi$ to the form $\psi=a(r) e^{i\varphi(r)}$,
while merely adding a constant to $A_t$.

Now on the one hand, the radial current component is
  \begin{equation}
J_r= \imath(\psi^*\partial_r\psi - \psi\partial_r\psi^*) =
 -2 a^2 \partial_r\varphi
\label{current}
\end{equation}
On the other hand, conservation of charge together with the symmetries implies
that
\begin{equation}
	J_r\,e^{-\lambda}\sqrt{-g}={\rm const.}
\end{equation}
The constant here must vanish; otherwise charge would leak out continually
to infinity.  It follows from Eq.(\ref{current}) that $\varphi$ cannot depend
on
$r$, except possibly for jumps at nodes where $a=0$.
However, an arbitrary jump of $\varphi$ through a node causes an unacceptable
discontinuity in $\psi,_r$ unless the jump is by an odd multiple of $\pi$ in
which case its effect on $\psi,_r$ is cancelled by the change in sign of
$a,_r$ at the node.  Accordingly, in all that follows we regard $\varphi$ as
strictly constant (and thus irrelevant) at the cost of allowing $a(r)$ to
change sign.
Thus in our problem $\psi$, henceforth denoted by $a$, is real while
$A_\mu$ reduces to a static temporal component.

In what follows it will be convenient to look at the differences
\begin{eqnarray}
	T_t^t-T_\phi^\phi &=&
	\xi e^{-\lambda} {(2/r-\nu')aa,_r \over  1-8\pi G\xi a^2}
	-{Q^2/(4\pi r^4)+q^2e^{-\nu} A_t^2a^2\over  1-8\pi G\xi a^2}
\label{Ttt-Tphph} \\
        T_t^t-T_r^r &=& e^{-\lambda}{
       (2\xi-1) a,_r^2 - \xi (\nu+\lambda)' aa,_r + 2\xi aa,_{rr}
	    \over {1-8\pi G \xi a^2}}-{q^2e^{-\nu} A_t^2a^2
		\over  1-8\pi G\xi a^2}
\label{Ttt-Trr}
\end{eqnarray}
(here and henceforth\ $'\,\equiv \partial/\partial r$),
as well as at the negative of the energy density
\begin{eqnarray}
	T_t^t&=&e^{-\lambda}{(2\xi-1/2){a,_r}^2
		+2\xi aa,_{rr}+\xi (4/r-\lambda') aa,_r\over 1-8\pi G\xi a^2}
\nonumber \\
	&-&{1\over 2}{V+Q^2/(4\pi r^4)+ q^2 A_t^2a^2e^{-\nu}\over  1-8\pi
G\xi a^2}
\label{Ttt}	
\end{eqnarray}

\subsection{The  Scalar--Maxwell Field Equations}
\label{THaMFE}

The field equation for $a$ that follows from Eq.(\ref{L}) is
\begin{equation}	
		D_\mu D^\mu a -  (\xi R + \dot V)a = 0.
\label{field_eq}
\end{equation}
where $\dot V\equiv\partial V(a^2)/\partial a^2$.  After substitution of the
metric functions and simplification we get \begin{equation}
	a,_{rr} + (1/2)(4/r+\nu'-\lambda') a,_r
                   -(\xi R + \dot V-q^2 e^{-\nu} A_t^2) e^\lambda a = 0,
\label{scalar}
\end{equation}

Finally the temporal component of the Maxwell equations
\begin{equation}
	{F^{\mu \nu}}_{;\mu} = 4\pi J^\nu
\end{equation}
takes the form
\begin{equation}
	A_t,_{rr} + (1/2)(4/r - \nu'-\lambda') A_t,_r
                  - 4 \pi q^2 a^2 e^\lambda A_t = 0
\label{maxwell_eq}
\end{equation}
Note that when $q=0$ the equations for $A_t$ and for the scalar field decouple
so that we can consider the two fields separately.

\subsection{Finiteness of $T_\mu^\nu$ and the Causality Restriction}
\label{TFoTaC}

There are two types of restrictions on the total energy--momentum tensor
which must be obeyed everywhere in the black hole exterior and horizon in
order for a solution to be physically acceptable: boundedness of the mixed
components
$T_\mu^\nu$, and the causality restriction.

Staticity and spherical symmetry imply that the only nonvanishing mixed
components of $T_\mu^\nu$ are $T_t^t$, $T_r^r$, and
$T_\theta^\theta=T_\phi^\phi$. Thus $T_{\alpha \beta} T^{\alpha
\beta}=(T_t^t)^2+(T_r^r)^2 + (T_\theta^\theta)^2 +(T_\phi^\phi)^2$.  But
$T_{\alpha \beta} T^{\alpha\beta}$ is a physical invariant and must thus be
bounded everywhere, including at the horizon: any divergence would imply
divergence of the curvature invariant $G_{\mu\nu} G^{\mu\nu}$.  It follows
that the {\it mixed\/} components $T_t^t$, $T_r^r$, and
$T_\theta^\theta=T_\phi^\phi$ are all bounded everywhere including at the
horizon (this is no longer true for a component like  $T_{rr}$).

Along with finiteness of $T_\mu^\nu$ in general we should cite here an
important result to be obtained in  Secs.\ref{Nonextremal} and
\ref{extremal}; it has also been noticed by Achucarro, Gregory and Kuijken
\cite{Achu} and by N\'u\~nez, Quevedo and Sudarsky \cite{Nunez}.  For any
static and spherical black hole and for any matter content, \begin{equation}
T_t^t(r_H)=T_r^r(r_H) \label{T=T} \end{equation}

Now consider the Poynting vector according to a physical observer with
4-velocity $u^\nu$
\begin{equation}
	j^\mu=-T_\mu ^\nu u^\mu
\end{equation}
with $u^\mu u_\mu =- 1$, and the associated energy density
\begin{equation}
\varepsilon\equiv T_{\mu\nu}u^\mu u^\nu
\end{equation}
If $\varepsilon>0$ then $j^\mu$ should be a non--spacelike four-vector, for
in this case $j^\mu$ defines a future--directed 4-velocity (with positive
time component $\propto \varepsilon$), and on grounds of causality this
``velocity of transfer of energy'' should not be superluminal.  If
$\varepsilon<0$ the Poynting vector points into the past.   We still expect
that $j^\mu$ should be non--spacelike because the flow of negative energy
can be interpreted as flow of positive energy in the opposite space direction
from that demarcarted by $j^\mu$.  In other words $j^\mu$  should, in this
case, point into the past lightcone.  Hence, for any observer we must have
\begin{equation}
	T_\mu^\nu u_\nu T^\mu_\sigma u^\sigma \leq 0
\label{causality}
\end{equation}

Now suppose that our observer moves in any way in the equatorial plane
$\theta=0$ in the hole's exterior,  Eq.(\ref{causality}) becomes:
\begin{equation}
	u^t u_t (T^t_t)^2 + u^r u_r (T^r_r)^2 + u^\phi u_\phi (T^\phi_\phi)^2
\leq 0
\label{special}
\end{equation}
Substituting here $u^t u_t$ from the normalization of the velocity
\begin{equation}
	u^t u_t + u^r u_r + u^\phi u_\phi= -1
\label{four-velocity}
\end{equation}
and rearranging the inequality gives
\begin{equation}
(T^t_t)^2 \geq{ u_r u^r (T^r_r)^2 + u_\phi u^\phi (T^\phi_\phi)^2\over 1 +
u^r u_r + u^\phi u_\phi}
\label{intermediate}
\end{equation}
In light of the positivity of $u^r u_r$ and $u^\phi u_\phi$, it follows that
inequality (\ref{special}) or (\ref{causality}) can be true for any velocity
if only if
\begin{equation}
\left| T^\theta_\theta\right|=\left| T^\phi_\phi\right|\leq\left|
T^t_t\right|\geq \left| T^r_r\right| \label{Ttt_dominant}
\end{equation}

The energy conditions (\ref{Ttt_dominant}) have been discussed by Hawking
and Ellis \cite{Hawk} who, however, considered them only for the positive
energy density case.  When dealing with nonminimal coupling to gravity,
negative energy density is not excluded.  In the Appendix we prove that
either the energy conditions  (\ref{Ttt_dominant}), or the causality
condition (\ref{causality}) for all observers, are equivalent to consensus of
all observers as to the sign of the energy density.  From all this it is
clear that the energy conditions (\ref{Ttt_dominant}) are a must for a
nonpathological solution of the field equations, and  henceforth we assume
them to hold.

The energy conditions (\ref{Ttt_dominant}) can also be stated as
\begin{equation}
{\rm Sgn}(T^t_t) = {\rm Sgn}(T^t_t-T^r_r) ={\rm
Sgn}(T^t_t-T^\phi_\phi)
\label{signs}
\end{equation}
We stress that no assumption is made here about the sign of the energy
density $-T^t_t$, so that these inequalities are more broadly valid than the
{\it weak\/} energy condition $T^r_r-T^t_t\geq 0$ which is sometimes invoked.

\section{MINIMALLY COUPLED NEUTRAL SCALAR FIELD WITH NONQUADRATIC ACTION}
\label{GtCBH}

There exists a no-hair theorem for black holes which rules out hair in
the form of a minimally coupled (to gravity), real multiplet scalar field for
any asymptotically flat, static, spherically symmetric neutral black hole
\cite{Bek95}. The field is assumed to  bear positive energy, but its field
Lagrangian need not be quadratic in the field derivatives.  Here we generalize
that theorem to charged black holes \cite{Mayo}, not only for its intrinsic
interest, but for use in our later theorems for nonminimally coupled fields.

Consider the action for real scalar fields, $\psi, \chi, \cdots$, accompanied
by an electromagnetic field
 \begin{equation}
	S_{\psi,\chi,\cdots}=-\int\left(
	{\cal E}({\cal I},{\cal J},{\cal K},\cdots\,\psi, \chi, \cdots)
		+ {1\over 16\pi}F_{\alpha\beta} F^{\alpha\beta}\right)
\sqrt{-g}d^4x
\label
{nonquadratic}
\end{equation}
Here ${\cal E}$ is a function (which for static fields turns out to be
identical to the energy density), and ${\cal I}\equiv
g^{\alpha\beta}\psi,_\alpha\psi,_\beta$, ${\cal J}\equiv
g^{\alpha\beta}\chi,_\alpha\chi,_\beta$ and ${\cal K}\equiv
g^{\alpha\beta}\chi,_\alpha \psi,_\beta$ are examples of the invariants
that can be formed from first derivatives of the scalar fields. We do not
assume that the kinetic part of the scalar's Lagrangian density can be
separated out, nor that it is a quadratic form in first derivatives.

Assume the existence of a spherically symmetric static black hole solution
with the said scalar fields as hair.  Because the scalar fields are assumed
decoupled from the electromagnetic field, the  energy--momentum tensor of the
scalar fields is conserved separately.  From the radial component of the
conservation law  ${T^{(sc)}}_\mu^\nu{}_{;\nu}=0$ together with the result
${T^{(sc)}}_t^t={T^{(sc)}}_\phi^\phi$ which follows from the form of
$S_{\psi,\chi,\cdots}$ and the symmetries, one obtains, as in \cite{Bek95},
the results  \begin{eqnarray}
	{T^{(sc)}}_r^r&=&-{e^{-\nu/2}\over r^2}
			\int_{r_H}^r \left(r^2e^{\nu/2}\right)'\rho dr
\label{iii} \\
	\left({T^{(sc)}}_r^r\right)'&=&
		-{e^{-\nu/2}\over r^2}\left(r^2 e^{\nu/2}\right)'
		\left(\rho + {T^{(sc)}}_r^r\right)
\label{iv}
\end{eqnarray}	
Here $\rho={\cal E}= -{T^{(sc)}}_t^t$ is the (assumed positive) energy
density of the scalar fields. In order for the mass to be finite, we shall
require that asymptotically $\rho=O(r^{-3})$.   Now the positivity of $\rho$,
the relation (\ref{Tem}) and the causality restriction (\ref{signs}) for the
overall $T_\mu^\nu$ tell us that $\rho+{T^{(sc)}}_r^r>0$.  Since $e^\nu$
vanishes at $r=r_H$ (see Sec.\ref{general} below) and must be positive for
$r>r_H$, $r^2e^{\nu/2}$ must grow with $r$ at least sufficiently near the
horizon. It is  then immediately obvious from Eqs.(\ref{iii}) and (\ref{iv})
that sufficiently near the horizon ${T^{(sc)}}^r_r$ and
$\left({T^{(sc)}}_r^r\right)'$ are both negative.

Asymptotic flatness considerations together with  Eqs.(\ref{iii}), (\ref{iv})
 tell us that as $r\rightarrow\infty$, ${T^{(sc)}}^r_r=O(r^{-2})$ and
$\left({T^{(sc)}}^r_r\right)'<0$. From these follows that asymptotically
${T^{(sc)}}^r_r > 0$ so that ${T^{(sc)}}^r_r$ must switch sign at some finite
point $r=r_c$.  Hence we infer that in some intermediate interval $[r_a,r_b]$,
$\left({T^{(sc)}}^r_r\right)'>0$ and the point where ${T^{(sc)}}^r_r$ changes
sign is $r_c\in[r_a,r_b]$.  ${T^{(sc)}}^r_r$ is positive in $[r_c,r_b]$.

Now it turns out that this last conclusion clashes with Einstein's equations.
The relevant one are  \begin{eqnarray}
	e^{-\lambda}\left({1\over r^2}-{\lambda '\over r}\right)-{1\over r^2}
		&=&8\pi GT_t^t=-8\pi G\left(\rho+{Q^2\over 8\pi r^4}\right)
\label{i}\cr
	e^{-\lambda}\left({\nu '\over r}+{1\over r^2}\right)-{1\over r^2}
		&=&8\pi G T_r^r=8\pi G\left({T^{(sc)}}_r^r-{Q^2
\over 8\pi r^4}\right)
\label{ii}\cr
\end{eqnarray}
where $Q$, a constant in the present section as well as in
Sec.\ref{xinontrivial}, is the total charge of the black hole.   Sometimes
the difference of these comes in handy:
\begin{equation}
	e^{- \lambda} (\nu ' + \lambda ') = -8 \pi G (T_t^t - T_r^r) r.
\label{eq.rr-eq.tt}
\end{equation}
Now by integrating Eq.(\ref{i}) out from the horizon  radius, $r_H$, and
solving for $e^{\lambda}$ we obtain \begin{equation}
	e^{-\lambda}=1-{r_H\over r}-{8\pi G\over r}
	\int_{r_H}^r \left(\rho+{Q^2\over 8\pi r^4}\right) r^2 dr
\label{v}
\end{equation}
where the integration constant has been set so that $e^{-\lambda}$ vanishes at
 $r_H$.  It follows from Eq.(\ref{v}) that $e^\lambda\geq 1$ throughout the
black hole exterior.

Consider now the second Einstein equation. Since $e^{-\lambda(r_H)}=0$, but
$\nu'(r)$ may diverge positively at $r=r_H$ (see Sec.\ref{general} below), we
can write it as
\begin{equation}
	8 \pi G {T^{(sc)}}^r_r(r_H) -{G Q^2\over {r_H}^4}
		\geq -{1\over {r_H}^2}
\end{equation}
Because in the proximity of the horizon ${T^{(sc)}}^r_r<0$, we can infer that
\begin{equation}
	GQ^2\leq{r_H}^2
\end{equation}
so that trivially
\begin{equation}
{-GQ^2\over 2r^3} +{1\over 2r}>0
\label{ineq}
\end{equation}

Now rewriting Eq.(\ref{ii}) we infer from Eq.(\ref{ineq}) that in a region
where ${T^{(sc)}}^r_r>0$ \begin{equation}
	{e^{-\nu/2}\over 2}\left(r^2 e^{\nu/2}\right)'=
\left(4 \pi G r {T^{(sc)}}_r^r
-{GQ^2\over 2r^3} + {1\over 2r}\right)e^\lambda + {3\over 2r}>0
\end{equation}
We found that in $[r_c,r_b]$, ${T^{(sc)}}_r^r>0$.  Thus
$(e^{-\nu/2}/2)\left(r^2 e^{\nu/2}\right)'>0$ there.  According to
Eq.(\ref{iv}) this means that $\left({T^{(sc)}}_r^r\right)'<0$ throughout
$[r_c,r_b]$. However, we determined that  $\left({T^{(sc)}}_r^r\right)'>0$
throughout the encompassing interval $[r_a,r_b]$. Thus there is a
contradiction. The only way to resolve it is to accept that the the scalar
field component must be constant throughout the black hole exterior, taking
values such that all components of ${T^{(sc)}}_\nu^\mu$ vanish identically.
Such values must exist in order that the trivial solution of the scalar field
equation be possible in free empty space. It is this solution which served
implicitly as an asymptotic boundary condition in our argument.

Thus {\it the unique asymptotically flat, static, spherically symmetric
black hole solution of the action (\ref{nonquadratic}) is the
Reissner-Nordstr\"{o}m black hole with no scalar hair.\/}

\section{NONMINIMALLY COUPLED ($\xi< 0$ or  $\xi\geq {1\over 2}$) NEUTRAL
SCALAR FIELD WITH SELF--INTERACTION}
\label{xinontrivial}
\subsection{Case $\xi<0$}
\label{map}
We now consider hair described by action (\ref{L}) with $q=0$
and a potential restricted by $V\geq 0$ for a black hole which may or may not
be charged.  In order to prove that the field can only  be in a trivial
configuration, we shall use a conformal map to show that in a new metric the
action is equivalent to that considered in Sec.\ref{GtCBH}.   This approach
has also been used by Saa \cite{Saa}, who also started from the theorem
discussed in Sec.\ref{GtCBH} in its neutral black hole version \cite{Bek95}.

With the proposed solution for the black hole with nonminimally coupled and
neutral scalar field hair, $\{a, g_{\mu\nu}\}$, we construct the map
\begin{eqnarray}
	g_{\mu\nu} &\rightarrow& \bar{g}_{\mu\nu}\equiv g_{\mu\nu}\Omega
\label{map1}
\nonumber \\
	\Omega &\equiv& 1-8 \pi G \xi a^2
\end{eqnarray}
For $\bar{g}_{\mu\nu}$ to be nondegenerate and of like signature to
$g_{\mu\nu}$, $\Omega$ must be strictly positive and bounded in $r\in [r_H,
\infty)$.  Obviously $\Omega>0$ here.  Saa leaves pending the question of
boundedness of the conformal factor.  We now {\it prove\/} that $\Omega$ is
bounded in the black hole exterior both for $\xi<0$ and for ${1\over 2}\leq
\xi$.

$\Omega $ can blow up only where $a$ does so.  However, at such a point $r_c$,
$a,_r$ and $a,_{rr}$ would be even more singular than $a$.  Specifically,
$a,_r^2/a^2$ and $a,_{rr}/a$ should both behave like $(r-r_c)^{-2}$.  We see
from Eqs.(\ref{Ttt-Tphph}) and (\ref{Ttt}) that if $r_c> r_H$,  the physical
components of the energy--momentum tensor definitely diverge at $r_c$, which
is unphysical (Sec.\ref{TFoTaC}).  If $r_c=r_H$, the factor $e^{-\lambda}$,
ameliorates the divergence.   According to Eq.(\ref{behavior_horizon}) for
a nonextremal black hole $e^{-\lambda} \sim (r-r_H)$; this is not enough to
cancel the divergence.  By contrast, for an extremal black hole
$e^{-\lambda} \sim (r-r_H)^2$  so it would seem that the divergence is
quenched. But since $a^2\rightarrow\infty$ for $r$ approaching $r_H$ from the
right, it is evident that $aa,_r$ and $aa,_{rr}>0$ near the singularity.
We thus see from Eq.(\ref{Ttt-Trr}) in its variant form (4.3)
that for $r\rightarrow r_H$, $T_t^t-T_r^r$ has a {\it
positive\/} definite limit for either $\xi<0$ or $\xi\geq {1\over 2}$.
However, Eq.(\ref{T=T}) tells us that  $T_t^t-T_r^r$ must vanish at
a regular spherical horizon.  Thus for all black holes and for  $\xi<0$ or
$\xi\geq {1\over 2}$, $\Omega $ can blow up in $r\in [r_H, \infty)$ only for
physically unacceptable solutions.  Additionally, if $a$ were to diverge for
$r\rightarrow\infty$, the various $T_\mu{}^\nu$ would blow assymptotically.
We conclude that $\Omega<\infty$ everywhere outside the black hole both for
$\xi<0$ and for $\xi\geq 1/2$.  And since $\Omega $ cannot vanish for
$\xi<0$ we see that the map is regular for physically acceptable
black holes and $\xi<0$.

Under the map the action (\ref{L}) together with the Hilbert-Einstein action
is  transformed into
\begin{eqnarray}
	S&=&{1\over 16\pi
G}\int{\bar{R}\sqrt{-\bar{g}}d^4x}-{1\over 2}\int{\left(
(1+f)\bar{g}^{\alpha\beta}a,_\alpha a,_\beta + \bar{V} +{1\over 8 \pi}
\bar{g}^{\alpha\gamma}\bar{g}^{\beta\delta} F_{\alpha \beta} F_{\gamma \delta}
		\right)\sqrt{-\bar{g}}d^4x}
\nonumber \\	
	 f&\equiv&48 \pi G \xi^2 a^2 (1-8 \pi G \xi a^2)^{-2}
\nonumber \\
	\bar{V}&\equiv&V(1-8 \pi G \xi a^2)^{-2}
\end{eqnarray}
The transformed action is of the form (\ref{nonquadratic}).  It is
easily checked that in the static situation the field bears positive energy
with respect to $\bar{g}_{\mu\nu}$, not least because of the assumed
positivity of $V(a^2)$.  It is also easily seen that the map leaves the
mixed components
$T_\mu^\nu$ unaffected.  Hence the finiteness of these, and the causality sign
relations (\ref{signs}), can be used in the new geometry.

There is one little
complication.  We know that $\Omega $ goes to some finite positive value
at infinity (an oscillatory behavior is excluded by the argument to be given
presently that $\Omega$ determines the effective gravitational constant, which
 cannot oscillate spatially in a physical solution).  Since this value is not
necessarily  unity, the asymptotically Minkowskian
metric $g_{\mu\nu}$ will be mapped into a not necessarily asymptotically
Minkowskian $\bar{g}_{\mu\nu}$.  But $\bar{g}_{\mu\nu}$ {\it is\/}
asymptotically flat.  One need only redefine globally the units of length and
time to make it of standard Minkowski form at infinity.  With this proviso we
may apply the theorem of Sec.\ref{GtCBH} to show that $a$ must be constant.

Thus {\it there exists no static spherically symmetric neutral or charged
black hole endowed with nontrivial hair in the form of a neutral scalar field
nonminimally coupled to gravity with $\xi<0$ and with a nonnegative
self--interaction potential.}

\subsection{Case $\xi\geq {1\over 2}$ with $Q\neq 0$}
\label{second}

Before starting that proof, we comment on the asymptotic value of $\Omega $.
This is determined by the value of $a$ for which $a,_r\rightarrow 0$ and
$a,_{rr}\rightarrow 0$ as $r\rightarrow \infty$ according to  the scalar
equation, Eq.(\ref{scalar}).
Since $q=0$ here and $R\rightarrow 0$ asymptotically, $a(\infty)$
is determined by $\dot V(a(\infty)^2)=0$.  Further, in order for the  energy
density $-T_t^t$ to vanish in the same limit (asymptotic flatness), we need,
according to Eq.(\ref{Ttt}), that $V(a^2)$ itself vanish where $\dot V(a^2)$
vanishes.  In addition, this common zero of $V$ and $\dot V$ must be such as
to make $\Omega >0$.  For otherwise the {\it effective\/} gravitational
constant would be negative far away from the black hole.  One way to see this
is to imagine adding to the background of the black hole solution with
energy--momentum tensor given by Eq.(\ref{TG}) a small positive mass.  In
Eqs.(\ref{Ttt-Tphph}) and (\ref{Ttt})   the additional energy--momentum
tensor would appear as  contributions to the numerators, with everything
divided by
$\Omega $.  In a region where $\Omega <0$ that mass would thus contribute to
the gravitational field as if it were negative.   This contribution will repel
a second particle of the same kind (treated as a test particle).  Thus
positive masses would repel each other gravitationally and the effective
gravitational constant $G_{\rm eff}=G(1-8\pi G\xi a^2)^{-1}$ would be
negative.  This is certainly unphysical if the region is far from the black
hole (it could be our neighborhood).   We conclude that a physically
reasonable black hole solution must have $\Omega >0$ asymptotically, which
requires that both $V(a^2)$ and $\dot V(a^2)$ have at least one common root
$a^2<(8\pi G\xi)^{-1}$.

We now proceed to prove  by contradiction that $\Omega$ cannot vanish in
$[r_{H}, \infty)$.  Suppose that there is a nontrivial physically reasonable
neutral black hole solution,
 for which $\Omega $ vanishes at some point
$r=\tilde{r}$ (if there are several points $\tilde{r}$, we focus on the {\it
rightmost\/} one).  It is obvious from Eq.(\ref{Ttt-Tphph}) that $a,_r\neq 0$
and $2/r-\nu'\neq 0$ at $\tilde{r}$ for if either vanished,
$T_t^t-T_\phi^\phi$ would necessarily diverge there contrary to the
requirements in Sec.\ref{TFoTaC}.  In fact $a^2,_r< 0$ at $r=\tilde{r}$
because $\Omega $ must be positive as $r\rightarrow\infty$ .

Now $a^2$ cannot have a minimum.  For at such point $r=\hat{r}$, $a,_r=0$ and
$aa,_{rr}>0$. Obviously $\hat{r}\neq \tilde{r}$ because we found $a,_r\neq 0$
at the latter.  But then according to Eqs.(\ref{Ttt-Tphph}) and
(\ref{Ttt-Trr}), $T_t^t-T_\phi^\phi$ and $T_t^t-T_r^r$ will have  opposite
signs at  $r=\hat{r}$ (we assume $\xi>{1\over 2}$).  But this contradicts the
causality restriction (\ref{signs}).  Thus in our solution $a(r)^2$ must be
monotonically decreasing.

It follows that near infinity we must have $aa,_r<0$ and $aa,_{rr}>0$.  From
asymptotic flatness it follows that $\nu '\sim 1/r^2$ for sufficiently large
$r$.   Hence by Eq.(\ref{Ttt-Tphph}), $T_t^t-T_\phi^\phi<0$ asymptotically.
By causality [Eq.(\ref{signs})] $T_t^t-T_r^r$  must then be negative for
large $r$.  This condition together with Eq.(\ref{eq.rr-eq.tt}) tells us that
asymptotically $(\nu+\lambda)'>0$.  Substituting all these in
Eq.(\ref{Ttt-Trr}) we find that $T_t^t-T_r^r>0$ for large $r$.  But this
contradicts our previous conclusion.  Our supposition that $\Omega $ vanishes
somewhere is thus rebutted, at least for $\xi\geq {1\over 2}$, $q=0$ and
$Q\neq 0$.

Recalling from Sec.\ref{map} that $\Omega$ cannot blow up in the black hole
exterior, we see that the map used there is equally valid in the present case.
Thus by the same logic as used in Sec.\ref{map}, hair is excluded in the
present case.

\subsection{Case  $\xi\geq {1\over 2}$ with $Q=0$} \label{first}

The vanishing of $Q$ compromises our proof in Sec. \ref{second} that $a^2$
has no minimum.  We thus adopt here a new strategy unrelated to the map
(\ref{map1}).  Again the proof of this claim proceeds by contradiction.  We
assume there is a
nontrivial physically reasonable neutral black hole solution with $Q= 0$.

Let us first eliminate $\nu'+\mu'$ from Eq.(\ref{Ttt-Trr}) with the help of
Eq.(\ref{eq.rr-eq.tt}) to get
\begin{equation}
T_t^t-T_r^r={e^{-\lambda}[(2\xi-1)a_r^2+2\xi a a,_{rr}]\over
1-8\pi G\xi a^2-8\pi G\xi r aa,_r}
\label{newTtt-Trr}
\end{equation}
Obviously as $r\rightarrow\infty$, $aa(r),_r$ must fall off faster than
$r^{-1}$, so that the denominator here is asymptotically positive by the
positivity of the asymptotic gravitational constant.

Now suppose that asymptotically $a^2$ decreases, which means that  $a a,_r<0$
and $a a,_{rr}>0$.  It is then plain from Eq.(\ref{Ttt-Tphph}) that
$T_t^t-T_\phi^\phi<0$ while from Eq.(\ref{newTtt-Trr}) it is clear that
$T_t^t-T_r^r>0$.  This would violate causality, and must be excluded.  Thus
suppose the opposite, that $a^2$ increases asymptotically so that $a^2,_r>0$
while $a^2,_{rr} <0$. Rewriting Eq.(\ref{newTtt-Trr}) in the form
\begin{equation}
T_t^t-T_r^r={e^{-\lambda}[\xi a^2,_{rr}-a,_r^2]\over
1-8\pi G\xi a^2-4\pi G\xi r a^2,_r}
\label{newnewTtt-Trr}
\end{equation}
we see that now $T_t^t-T_\phi^\phi>0$ while $T_t^t-T_r^r<0$.  This new
possibility is thus also ruled out by causality.  Likewise, were $a^2$ to
oscillate indefinitely as $r\rightarrow\infty$, a similar clash would
ensue over part of each cycle.  We must thus conclude that $a$ is strictly
constant for $r$ greater than some finite but large $r_*$.  A Taylor expansion
of $a(r)$ about a point to the right of $r_*$ must obviously sum up to the
asymptotic value $a(\infty)$.  Now the differential equation for $a$, Eq.
(\ref{scalar}) has singular points only at $r=r_{H}$ and $r=\infty$
($R$ must be bounded in the black hole exterior while we assume that
$V$ is a regular function).  Thus the series must converge to the correct
$a(r)$ all the way to the horizon and $a\equiv $ const. so that
there is no hair.

Summarizing this and the last section, {\it there exists no static spherically
 symmetric neutral or charged black hole endowed with nontrivial hair in the
form of a neutral scalar field nonminimally coupled to gravity with $\xi\geq
1/2$ and with a nonnegative and regular self--interaction potential.}

\section{NONMINIMALLY COUPLED (ANY $\xi$) CHARGED SCALAR FIELD
WITH SELF--INTERACTION}
\label{ChargedField}

Next we consider charged scalar hair, possibly nonminimally coupled to
gravity (any $\xi$) and with a  positive semidefinite self--interaction
potential assumed to be a regular function of its argument $a^2$.   We shall
here invoke a new strategy, namely looking at the analytic behavior of
various quantities in the horizon's vicinity, as dictated by the very nature
of the horizon.  The following two subsections contain general conclusions
about the horizon and its neighborhood which are independent of the matter
content of the black hole exterior, first in general and then for nonextremal
black holes.  These are extended to extremal black holes in Sec.\ref{last}.
In this section $Q(r)$ denotes the charge of black hole plus scalar field up
to radial coordinate $r$.

\subsection{General Properties of a Spherical Static Event Horizon}
\label{Horizon}
\label{general}

We return to  Eq.(\ref{v}) written as
\begin{equation}
	e^{-\lambda} =  1 - {r_H\over r} + {8 \pi G\over r} \int^r_{r_H}
T^t_t r^2 dr
\label{e_lambda}
\end{equation}
As anticipated already, the point $r=r_H$ where  $e^{-\lambda}$ vanishes is
to be interpreted as the location of the horizon. To see why define a family
of spherical hypersurfaces by the conditions $\{\forall t; f(r)=\mbox{\rm
const.}\}$ with $f$ monotonic.  Each value of the constant labels a different
surface. The normal to each such hypersurface is    \begin{equation}
	n_\mu = {\partial f\over \partial x^\mu} = (0,1,0,0) f'.
\end{equation}
Hence
\begin{equation}
	n_\mu n^\mu = e^{-\lambda}(f')^2
\end{equation}
which vanishes only for $r=r_H$.  This must thus be location of the horizon
which is defined as a null surface (hence null normal).

Proceeding with the argument, assume  that $e^\nu$ vanishes at some
point $\bar{r}$. Then $\nu\rightarrow -\infty$ and $\nu'\rightarrow \infty$ as
$r
\rightarrow \bar{r}$ from the right. It is then obvious from Eq.(\ref{ii})
that $e^{-\lambda}$ must vanish as $r\rightarrow \bar{r}$ since $T_r^r$ must
be bounded.   But since $e^{-\lambda}$ vanishes only for $r=r_H$, we see that
$\bar{r}=r_H$: $e^{-\lambda}$  vanishes wherever $e^\nu$ vanishes.  The
converse is also true: the horizon $r=r_H$ must always be an infinite redshift
surface with $e^\nu=0$.   For if $e^\nu$  were positive at $r=r_H$, then
according to the metric Eq.(\ref{metric}) the $t$ direction would be timelike
there, while the $\theta$ and $\phi$ directions would be, as always,
spacelike.  But since the horizon is a null surface, it must have a null
tangent direction, and this must obviously be the $t$ direction.    Thus it
is inconsistent to assume that $e^\nu\neq 0$ at $r=r_H$.

\subsection{Matter Independent Characterization of Nonextremal Event Horizon}
\label{Nonextremal}

Since $T_t^t$ must be bounded on the horizon, we may write the first
approximation (in Taylor's sense) for  $e^{-\lambda}$ near the horizon as:
\begin{equation}
	e^{-\lambda}=L(r-r_H)+O((r-r_H)^2);\qquad L\equiv
{1+8 \pi G T_t^t(r_H) r_H^2\over r_H} \label{e_lambda_first}
\end{equation}
Since $e^{-\lambda}$ must be non--negative outside the horizon, we  learn that
$L>0$, that is, {\it at every static spherically symmetric event horizon\/}
\begin{equation}
	- (8 \pi G r_H^2)^{-1} \leq T_t^t(r_H)
\label{Ttt_is_bounded}
\end{equation}
Note that the energy density at the horizon, if positive, is limited by the
very condition of regularity at the  horizon.  The inequality is saturated
for the extremal black hole; we consider this case in Sec.\ref{extremal}
below.

Under the assumption of asymptotic flatness, we can integrate
Eq.(\ref{eq.rr-eq.tt}) to get
\begin{equation}
	\nu +\lambda = 8 \pi G \int ^\infty _r r'(T_t^t-T_r^r) e^\lambda dr'
\label{nu+lambda}
\end{equation}
Here the $T^\nu_\mu$ are finite everywhere, including at the horizon, and
$T_t^t$ must vanish  asymptotically faster than $1/r^3$ in order for
$e^\lambda$ not to diverge at infinity [see Eq.(\ref{e_lambda})]. In view of
Eq.(\ref{Ttt_dominant}) the difference $T_t^t -T^r_r$  vanishes at least as
fast as $1/r^3$. We thus conclude that $\nu+\lambda$ is regular
everywhere, except possibly on the horizon.

Now in view of Eq.(\ref{e_lambda_first}) we get from  Eq.(\ref{nu+lambda})
\begin{equation}
	\nu+\lambda={\rm const.}-
		8\pi G r_H^2 {T^t_t(r_H)-T^r_r(r_H)\over
			      1+8 \pi G T^t_t(r_H) r_H^2} \ln (r-r_H) +O(r-r_H)
\label{nu+lambda_first}
\end{equation}
But Eq.(\ref{e_lambda_first}) informs us that
\begin{equation}
	\lambda = {\rm const.} - \ln (r-r_H)  + O(r-r_H)
\label{lambda}
\end{equation}
Thus
\begin{equation}
	\nu ={\rm const.} + \beta \ln (r-r_H) + O(r-r_H);\qquad \beta
\equiv {1+8 \pi G T^r_r(r_H) r_H^2\over 1+8 \pi G T^t_t(r_H) r_H^2}
\label{nu_first}
\end{equation}

The value of $\beta$ is restricted by the requirement that the scalar
curvature
\begin{equation}
	R=e^{-\lambda} \left(\nu ''+{1\over 2}\nu '^2
			   +{2\over r} (\nu ' -\lambda ')
	                   -{1\over 2}\nu ' \lambda '
			   +{2\over r^2} \right)-{2\over r^2}
\end{equation}
be bounded on the horizon (this is the same as boundedness of $T$).   If we
substitute here Eqs.(\ref{lambda}) and (\ref{nu_first}) we get
\begin{equation}
	R = -{2\over r_H^2} +L\ (r-r_H ) \times \left({1\over 2} {\beta(\beta-1)
\over (r-r_H)^2}+{2\over r_H}{\beta+1 \over (r-r_H)}+{2\over r_H^2} \right)
\label{u->R} \end{equation}
Obviously the terms in Eq.(\ref{u->R}) that diverge at the fastest rate must
cancel. Since we are considering a nonextremal black hole, $L>0$, so we are
left with the condition  \begin{equation}
	\beta (\beta -1)=0
\end{equation}
The alternative $\beta=0$ is excluded by the requirement (Sec.\ref{Horizon})
that $e^\nu=0$ at the horizon.  Thus necessarily $\beta = 1$.  We thus recover
Eq.(\ref{T=T}).  In addition, we learn that
\begin{equation}
	e^\nu = N (r-r_H) + O((r-r_H)^2)
\label{e_nu}
\end{equation}
where $N$ denotes a positive constant.

\subsection{$A_t$ is Bounded on the Horizon}
\label{At}

Our choice of gauge in Sec.\ref{TEMT} does not guarantee that
$A_t(\infty)=0$.  For
that same gauge transformation with $\Lambda= {\rm const.\ }\times t$ which
we used to make $\psi$ static adds a constant to $A_t$, and so may make
$A_t(\infty)\neq 0$.  To show this does not happen in a physically
acceptable solution, we assume otherwise and exhibit a contradiction.  Thus
suppose $A_t(\infty)\neq 0$ with $A_t,_r$ and $A_t,_{rr}$ vanishing
asymptotically.  Then it follows from Maxwell's Eq.(\ref{maxwell_eq}) that
$a(\infty)=0$ so that all derivatives of $a$ vanish asymptotically.  Putting
this fact together with the requirement from
asymptotic flatness that $T_t^t\rightarrow 0$ into Eq.(\ref{Ttt}), we see
that the potential must satisfy $V(0)=0$.  But the potential is
positive semidefinite so that  we must also require that $V'(0)=0$.

Turn now to the scalar equation Eq.(\ref{scalar}) and realize that because
of the asymptotic vanishing of $R$, the equation must reduce as $r\rightarrow
\infty$ to
\begin{equation}
a,_{rr}+2r^{-1} a,_r - q^2 A_t(\infty)^2 a = 0
\label{simple}
\end{equation}
with general solution
\begin{equation}
a = K r^{-1} \sin\left(qA_t(\infty) r + \chi\right)
\label{solu}
\end{equation}
where $K$ and $\chi$ are integration constants.  Although this $a$ falls off
asymptotically, it does so too slowly.  The electric charge density
it implies,
\begin{equation}
\rho \propto q^2 A_t a^2 \propto r^{-2} \sin\left(qA_t(\infty)r+\chi\right)
\label{charge}
\end{equation}
leads to a total charge which diverges asymptotically as $r$.  The
implication that the black hole is surrounded by a cloud with infinite
charge is clearly physically unacceptable.  We conclude that the assumption
$A_t(\infty)\neq 0$ is incompatible with a physically acceptable solution.
We shall thus assume henceforth that $A_t(\infty)=0$.

We shall now prove that $|A_t|$ is a monotonically decreasing function of $r$.
$F_{tr}$ must obviously vanish at spatial infinity. Consider the case that
$A_t$ is of one sign throughout and, with no
loss of generality, assume that $A_t$ is non--negative. Assume further that
$A_t$ has an extremum at some point $r=\hat{r}$ outside of the horizon.
But according to Eq.(\ref{maxwell_eq}), at any extremum $\mbox{\rm Sgn}
(A_t,_{rr})=\mbox{\rm Sgn} (A_t)$ so that an extremum must be a minimum.  On
the other hand, since $A_t$ vanishes asymptotically, it cannot have a minimum
without also having a maximum.   There is thus a contradiction which signals
the incorrectness of the assumption that there is an extremum.

When $A_t$ can change sign, assume with  no loss of generality that
$A_t$ changes from negative to positive with increasing $r$. In that case
$A_t$ would have to attain a positive maximum in order for $A_t \rightarrow
0$ as $r\rightarrow \infty$.  But the previous argument shows that $A_t$ is
forbidden positive maxima.  Thus $|A_t|$ cannot change sign.  It follows from
the preceding argument that $|A_t|$ must be monotonically decreasing in $r$.

Introduce now the set of  orthonormal differential forms
\begin{eqnarray}
	d\hat{t} &=& -e^{\nu \over 2} dt
\nonumber \\
	d\hat{r} &=& e^{\lambda \over 2} dr
\nonumber \\
	d\hat{\theta} &=& r d\theta
\nonumber \\
	d\hat{\phi} &=& r \sin \theta d\phi
\end{eqnarray}
The physical components of the Faraday tensor, $\hat{F}_{\mu\nu}$, are
related to the coordinates components by $\hat{F}_{\mu\nu} d\hat{\omega}^\mu
\wedge d\hat{\omega}^\nu = F_{\mu \nu} dx^\mu \wedge dx^\nu$, so that
\begin{equation}
	\hat{F}_{tr} e^{\nu+\lambda \over 2}	=F_{tr}
\end{equation}
The physical component $\hat{F}_{tr}$ must be finite.  From
Eqs.(\ref{e_lambda_first}) and (\ref{e_nu}) we see that $e^{\nu+\lambda \over
2}$  is bounded at the horizon.  Thus $A_{t},_{r}=-F_{tr}$ must be bounded at
the horizon. Integrating it once we obtain \begin{equation}
	A_t=\mbox{\rm const.} - {\rm const.} \times (r-r_H),
				\qquad
					  r \rightarrow r_H
\label{A_t}
\end{equation}
This completes our proof.

\subsection{Proof for Nonextremal Black Hole}
\label{a}
First consider the Maxwell equation (\ref{maxwell_eq}).  We know that
$A_t,_r$ must be bounded on the  horizon, so that even if $A_t,_{rr}$ diverges
there, it can only do so slower than  $(r-r_H)^{-1}$.   Now since $e^\lambda$
diverges as $(r-r_H)^{-1}$ while $(\nu+\lambda)'$ remains bounded, $a$ must
vanish on the horizon; otherwise, the last term in the equation would blow up
without being balanced.

We now look at the scalar equation (\ref{scalar}).  If the potential is
regular as assumed, $\dot V$ has to be bounded as $a\rightarrow 0$ at the
horizon.  The curvature $R$ is likewise bounded by assumption of a regular
horizon.  Therefore, according to Eqs.(\ref{e_nu}) and (\ref{A_t}), the last
term of the equation is dominated by the factor proportional to $q^2$.  It
follows from Eq.(\ref{e_lambda_first}) that near the horizon the scalar
equation has the limiting form  \begin{equation}
a,_{rr} + (r-r_H)^{-1} a,_r + (LN)^{-1} q^2  A_t(r_H)^2 (r-r_H)^{-2}
a = 0
\label{reduced}
\end{equation}
The two solutions of this Euler equation are
$(r-r_H)^{\pm\imath\alpha}$ with $\alpha\equiv qA_t(r_H)(NL)^{-1/2}$.
Combining them we get the general solution
\begin{equation} a(r)=B\sin \Phi; \qquad
\Phi\equiv \alpha \ln[(r-r_H)/D] \end{equation}
with $B$ and $D$ arbitrary constants.

Obviously for no choice of the constants does $a(r)$ vanish for $r\rightarrow
0$ as required. Not only that, but when we substitute this $a(r)$ into the
expressions (\ref{Ttt-Tphph}), (\ref{Ttt-Trr}) and (\ref{Ttt}) for the
components of $T_\mu^\nu$, every derivative of $a$ brings out a factor
$(r-r_H)^{-1}$, so that the expressions are singular at the horizon.  For
instance, from Eqs.(\ref{Ttt-Tphph}),  (\ref{Ttt-Trr}),
(\ref{e_lambda_first}) and (\ref{nu_first}) we get near the horizon
\begin{equation}
 T_r^r-T_\phi^\phi=-{L B^2\over r-r_H}\,
	{\alpha \sin \Phi(\xi \cos\Phi + \alpha \sin\Phi)
	     +O(r-r_H) \over 1-8\pi G \xi B^2\sin^2\Phi}
\label{temp}
\end{equation}
Obviously $T_r^r-T_\phi^\phi$ cannot remain bounded on the horizon as
required.  Thus
the solution with regular horizon we have been assuming is untenable.

In conclusion {\it there exists no non-extremal static and spherical charged
black hole endowed with hair in the form of a charged scalar field, whether
minimally or nonminimally coupled to gravity, and with a regular
positive semidefinite self--interaction potential.}

\subsection{Matter Independent Characterization of Extremal Event Horizon}
\label{extremal}

When inequality (\ref{Ttt_is_bounded}) is saturated, namely when
\begin{equation}
	T_t^t(r_H) = - ( 8 \pi G r_H^2)^{-1}
\label{saturation}
\end{equation}
we must continue the expansion of $e^{-\lambda}$ to second order:
\begin{equation}
	e^{-\lambda}={\cal L}(r-r_H)^2+O((r-r_H)^3);\qquad {\cal L}\equiv
{4\pi G
r_H^3T_t^t\,\acute{}\,(r_H) -1\over r_H^2}   \label{e_lambda_second}
\end{equation}
Because $e^{-\lambda}$ has to be positive for $r>r_H$,  ${\cal L}>0$ or
\begin{equation}
T_t^t\,\acute{}\,(r_H) > (4\pi G r_H^3)^{-1}
\label{T'}
\end{equation}
Of course Eq.(\ref{T=T}) must still hold since the saturated case is a
special member of the black hole family which can be reached continuously
from the main branch.  We note that  Eqs.(\ref{saturation}),
(\ref{e_lambda_second}) and (\ref{T=T}) are all satisfied at the extremal
Reissner--Nordstr\"om and BMBB black hole horizons.

Substituting these results in Einstein's equation (\ref{ii}), expanding
$T_r^r$ about its value at $r=r_H$, solving for $\nu'$,  and integrating we
have  \begin{equation}
 \nu = {\rm const.} + 2 \kappa \ln(r-r_H) +O(r-r_H); \qquad
\kappa={4\pi G r_H^3 T_r^r\,\acute{}\,(r_H) - 1\over 4\pi G r_H^3
T_t^t\,\acute{}\,(r_H) - 1}
\label{new_nu}
\end{equation}
We now show that causality restricts the possible values of $\kappa$.

Assume that  $T_t^t\,\acute{}\,(r_H) \neq T_r^r\,\acute{}\,(r_H)$.  Then in
light of Eq.(\ref{T=T}) we may expand near the horizon  \begin{equation}
T_t^t(r) - T_r^r(r) = (T_t^t\,\acute{}-T_r^r\,\acute{}\,)|_{r_H}(r-r_H) +
O((r-r_H)^2)
\label{difference}
\end{equation}
Because $T_t^t(r_H)<0$, $T_t^t(r)$ must be negative in a neighborhood of
the horizon.  The causality condition (\ref{signs}) then tells us that in that
same neighborhood, $T_t^t(r)<T_r^r(r)$. Then Eq.(\ref{difference}) implies
that  $T_r^r\,\acute{}\,(r_H) > T_t^t\,\acute{}\,(r_H)$. In light of
Eq.(\ref{T'}) this means that $\kappa>1$ in Eq.(\ref{new_nu}).

Thus the assumption $T_t^t\,\acute{}\,(r_H) \neq T_r^r\,\acute{}\,(r_H)$
implies that $e^\nu$ vanishes at the horizon faster than $(r-r_H)^2$
[presumably as $(r-r_H)^4$ if the metric coefficients are to avoid branch
points at the horizon and if the metric is not to change signature upon
traversal of the horizon].  However, there is nothing wrong with the
possibility $T_t^t\,\acute{}\,(r_H) = T_r^r\,\acute{}\,(r_H)$; it would simply
mean that the second order term in Eq.(\ref{difference}) is not allowed to
be positive.  In fact $T_t^t\,\acute{}\,(r_H) = T_r^r\,\acute{}\,(r_H)$, which
corresponds to $\kappa=1$, is attained at the extremal Reissner-Nordstr\"om
and BMBB horizons.  In view of all these facts we find it natural to define
extremal black holes as those characterized by Eqs.(\ref{saturation}),
(\ref{e_lambda_second}) and (\ref{T=T})  together with
\begin{equation}
e^\nu = {\cal N} (r-r_H)^2 + O((r-r_H)^3)  \label{e_nu_second}
\end{equation}
where ${\cal N}$ is a positive constant.  Higher order black holes with
$\kappa=2, 3,\cdots\ $ may not exist, just as third and higher order phase
transitions do not.

\subsection{Proof for Extremal Black Hole}
\label{last}
With the extremal black hole forms of the metric near the
horizon,Eqs.(\ref{e_lambda_second}) and (\ref{e_nu_second}), no
change transpires in the conclusions of Sec.\ref{At}, namely, the field $A_t$
must be monotonic in $r$, and from the regularity of the physical components
of
$F_{\mu\nu}$ one concludes that $A_t$ attains a bounded and nonvanishing value
at the horizon. Repeating the argument in Sec.\ref{a} with the new forms  of
the metric coefficients, one concludes that $a$ must vanish at the horizon
faster than $(r-r_H)^{1/2}$ in order for the Maxwell equation
(\ref{maxwell_eq}) to hold.

With this in mind let us look at the scalar equation (\ref{scalar}) in
the neighborhood of the horizon.  Recall that  $R$ and $\dot V(0)$ must
be bounded, so the corresponding terms are negligible compared with the $q^2$
term.  Substituting Eqs. (\ref{e_lambda_second})  and (\ref{e_nu_second}) and
retaining the leading contributions we get  \begin{equation}
	a,_{rr} + 2(r-r_H)^{-1} a,_r + ({\cal L N})^{-1} q^2  A_t(r_H)^2
(r-r_H)^{-4}
 a = 0
\end{equation}
which is to be contrasted with Eq.(\ref{reduced}).  In the variable
$u=(r-r_H)^{-1}$ we have \begin{equation}
	a,_{uu} + \bar\alpha^2 a = 0; \qquad \bar\alpha\equiv qA_t(r_H)({\cal L
N})^{-1/2} \end{equation}
with general solution
\begin{equation}
a(r) = {\cal B} \sin\bar\Phi; \qquad
\bar\Phi\equiv \bar\alpha (r-r_H)^{-1}+\zeta
\end{equation}
where ${\cal B}$ and $\zeta$ are integration constants.
For no choice of ${\cal B}$ and $\zeta$ does $a$ vanish for $r\rightarrow
r_H$ as required.  In addition, its very singular derivative leads, for
instance, to the expression
\begin{equation}
 T_r^r-T_\phi^\phi={L B^2\over(r-r_H)^2}\,
	{\alpha \sin \bar\Phi(2\xi \cos\bar\Phi - \alpha \sin\bar\Phi)
	     +O((r-r_H)^2) \over 1-8\pi G \xi B^2\sin^2\bar\Phi}
\end{equation}
which is incompatible with a regular horizon.  Thus the theorem stated at the
end of Sec.\ref{a} is extended to extremal black holes.

\section{Conclusions and Speculations}
\label{conclusions}

We have extended to charged static spherical black holes the exclusion of
hair in
the form of a neutral scalar multiplet with action which need not be
quadratic in the derivatives.  From this theorem we have excluded, for
charged or neutral static spherical black holes, hair in the form of a
neutral scalar field with standard kinetic action, positive  semidefinite
self--interaction potential and nonminimal coupling to gravity with $\xi<0$
and $\xi\geq {1\over 2}$. Finally, for charged static spherical black holes,
we have excluded hair in the form of a charged scalar field with standard
kinetic action, regular self--interaction potential, and nonminimal coupling
to gravity with any $\xi$.

Extension of the theorem excluding the neutral scalar field to the full
range $0<\xi<{1\over 2}$ is blocked by the existence of the BMBB black hole,
an extremal spherical black hole solution for the case $\xi={1\over 6}$ with
no self--interaction.   Xanthopoulos and Zannias \cite{Xanthopoulos,Zannias}
have shown that there are no more static black holes in this case,  even if
extremality or spherical symmetry are given up.  It may be that $\xi={1\over
6}$ is the unique value for which nonminimally coupled scalar black hole hair
appears.  In that case it should not be prohibitively difficult to produce a
single theorem proving this.  But if there exists a whole family of black
holes with nonminimally coupled hair within the domain $0<\xi<{1\over 2}$, of
which the BMBB black hole is just one example,  it would seem that at least
two theorems involving different approaches would be needed to exclude the
unoccupied hair parameter space on both sides of the putative family.

It seems unlikely that slightly aspherical charged black holes with
self--interacting neutral or charged scalar hair exist.  For one
would expect any such family to be governed by a parameter quantifying the
departure from spherical symmetry.  This parameter should reach the
spherical black hole.  Yet the spherical example is rigorously ruled out by
our theorems.  This heuristic argument obviously cannot be applied to very
aspherical black holes, or to those which show a topological distinction from
the spherical one.  Such is the case of the AGK black hole, a charged black
hole with minimally coupled self--interacting (Higgs) scalar hair in the form
of a local cosmic string which transfixes the black hole.  Strictly
speaking, our third theorem does not rule out such a solution because of its
lack of  spherical symmetry and asymptotic flatness.  But it is really the
distinct topology of the scalar field phase with its multiple connectivity
around the string which makes our proof far from relevant.

One can speculate on more complicated situations.  Suppose a black hole
forms with two local Higgs strings through it.  The situation would
seem unstable. Strings with the same sense of winding of the phase repel each
other, so the two strings will become antiparallel and approach.  If the
winding numbers were originally equal in absolute value, the strings will
anhilate with the Higgs phase topology becoming simple.  The configuration
will then relax.  But by our third theorem the endpoint cannot be a spherical
black hole with Higgs hair.  With due caution we infer that the black hole
will swallow part of the field and jettison the rest, so that we end up with
a Reissner--Nordstr\"om hole.  By extension we may surmise that if a black
hole is transfixed by an even number of unit winding--number strings, it will
end up with no scalar field, whereas if it has an odd number, it will end up
in the AGK configuration.

{\bf ACKNOWLEDGMENTS} J.D.B. thanks Norbert Straumann, Mikhail Volkov and
George Lavrelashvili for useful comments and Markus Heusler and Daniel
Sudarsky for correspondence.

\appendix
\section{The Energy Conditions}

At a given spacetime event consider the eigenvalue problem
\begin{equation}
T_\mu^\nu w^\mu = \sigma w^\nu
\label{eigenvalue}
\end{equation}
Because $T_\mu^\nu$ is a $4\times 4$ matrix, there must be four
distinct eigenvectors $w^\mu$.  Obviously
\begin{equation}
0\equiv w^{(2)}_\nu T_\mu^\nu w^{(1)\mu} - w^{(1)}_\nu T_\mu^\nu w^{(2)\mu}
= (\sigma^{(1)}-\sigma^{(2)})w_{(2)\nu} w^{(1)\nu}
\end{equation}
Hence for distinct eigenvalues the eigenvectors are orthogonal with respect
to the spacetime metric (for degenerate eigenvalues they can be made
orthogonal by the Schmidt procedure).  We gloss over the possibility that some
eigenvectors may be null (radiative solutions).  Thus one must be timelike;
call it $w^{(0)\mu}$ and normalize so that $w^{(0)\mu}w^{(0)}_\mu=-1$.  The
other three must be spacelike; call them $\{w^{(1)\mu}, w^{(2)\mu},
w^{(3)\mu}\}$ and normalize them so that $w^{(1)\mu}w^{(1)}_\mu=+1$, {\it
etc.\/}

The four eigenvectors obviously furnish a basis for writing any 4-vector, in
particular the 4-velocity of an observer:
\begin{equation}
u^\mu=c^{(0)}w^{(0)\mu}+\Sigma_i^3 c^{(i)}w^{(i)\mu}
\label{velocity}
\end{equation}
where obviously
\begin{equation}
(c^{(0)})^2 = \Sigma_i^3 (c^{(i)})^2 + 1
\label{normalization}
\end{equation}
in order to satisfy $u_\mu u^\mu=-1$.  The various choices of $\{c^{(i)}\}$
label all possible observers at a given event.

Now use Eqs.(\ref{eigenvalue}), (\ref{velocity}) and (\ref{normalization})
and the normalizations to reexpress (see Sec.\ref{TFoTaC})
\begin{equation}
j^\mu j_\mu = - (\sigma^{0})^2 -\Sigma_i (c^{(i)})^2
[(\sigma^{(0)})^2-(\sigma^{(i)})^2]
\end{equation}
and
\begin{equation}
\varepsilon\equiv T_{\mu\nu} u^\mu u^\nu = - \sigma^{0} -\Sigma_i
(c^{(i)})^2 (\sigma^{(0)}-\sigma^{(i)})
\end{equation}
We now see that the energy conditions
\begin{equation}
|\sigma^{(0)}|\geq \left\{|\sigma^{(i)}|\right\}
\end{equation}
are necessary and sufficient for $j^\mu j_\mu$ to be nonpositive
for all observers (all choices of $\{c^{(i)}\}$) and for the
energy density $\varepsilon$ to be of like sign (that of $-\sigma^{(0)}$) for
all observers.  Likewise  $j^\mu j_\mu\leq 0$ for all observers is a
necessary and sufficient condition for the energy conditions to be satisfied.
And consensus of all observers as to the sign of the energy density is
necessary and sufficient for the energy conditions to be satisfied and the
causality condition $j^\mu j_\mu\leq 0$ to hold for all observers.

In the static spherically symmetric situation considered in Sec.\ref{TFoTaC},
the $T_\mu^\nu$ is diagonal, so that $\sigma^{(0)}=T_t^t$,
$\sigma^{(1)}=T_r^r$,
$\sigma^{(2)}=\sigma^{(3)}=T_\theta^\theta=T_\phi^\phi$.
We thus recover the energy conditions (\ref{Ttt_dominant}).

\end{document}